\documentclass[preprint,amssymb,amsmath,aps,floatfix,prl]{revtex4}

\usepackage{graphicx}

\begin{document}

\title{Lorentz invariance violation and electromagnetic field in an intrinsically anisotropic spacetime}

\author{Zhe Chang$^{1,2}$\footnote{E-mail: changz@ihep.ac.cn}}
\author{Sai Wang$^{1}$\footnote{E-mail: wangsai@ihep.ac.cn}\footnote{E-mail: saiwangihep@gmail.com}
\footnote{Corresponding author at IHEP, CAS, 100049 Beijing, China}}
\affiliation{${}^1$\small{Institute of High Energy Physics\\Chinese Academy of Sciences, 100049 Beijing, China}\\
${}^2$\small{Theoretical Physics Center for Science Facilities\\ Chinese Academy of Sciences, 100049 Beijing, China}}

\begin{abstract}
Recently, Kostelecky [V.A. Kostelecky, Phys. Lett. B {\bf 701}, 137 (2011)] proposed that the spontaneous Lorentz invariance violation (sLIV) is related to Finsler geometry. Finsler spacetime is intrinsically anisotropic and induces naturally Lorentz invariance violation (LIV). In this paper, the electromagnetic field is investigated in locally Minkowski spacetime. The Lagrangian is presented explicitly for the electromagnetic field. It is compatible with the one in the standard model extension (SME). We show the Lorentz--violating Maxwell equations as well as the electromagnetic wave equation. The formal plane wave solution is obtained for the electromagnetic wave. The speed of light may depend on the direction of light and the lightcone may be enlarged or narrowed. The LIV effects could be viewed as influence from an anisotropic media on the electromagnetic wave. In addition, the birefringence of light will not emerge at the leading order in this model. A constraint on the spacetime anisotropy is obtained from observations on gamma--ray bursts (GRBs).
\end{abstract}

\maketitle

\section{1. Introduction}
At experimentally attainable energy scales, Einstein's special relativity (SR) is compatible with the present observations.
However, the SR is believed to be modified at higher energy scales,
such as the Planck scale which involves the effects of quantum gravity \cite{DSR01,DSR02,DSR03,DSR04,DSR05,spacetime foam01,spacetime foam02,spacetime foam03,spacetime foam04,VSR,Pavlopoulos,KosteleckyS001,SME01,SME02}.
The study on the string theory reveals that the Lorentz symmetry could be broken spontaneously
in the perturbative framework \cite{KosteleckyS001,SME01,SME02}.
The spontaneous Lorentz invariance violation (sLIV) involves nonzero vacuum expectation values of certain tensor fields.
It characterizes the anisotropy of spacetime since nonzero vacuum expectations of tensor fields are related to certain preferred directions.

To demonstrate spirit of the sLIV, we review shortly the spontaneous symmetry breaking in the electroweak theory.
The electroweak theory involves a Higgs field acquiring a nonzero vacuum expectation value, which leads to the mass terms of other particles.
Similarly, certain tensor fields acquire nonzero vacuum expectation values in the sLIV framework.
However, these expectation values take along the spacetime indices, which are different from the scalar one in the standard model (SM).
Therefore, the velocities of particles and fields may influence propagations and interactions, respectively.

Actually, the sLIV terms are added into the Lagrangian of fields by considering the gauge invariance, renormalizability, etc.
The vacuum expectation values of tensor fields become the coupling constants in the sLIV terms.
This approach to introduce the sLIV effects is called the standard model extension (SME) \cite{SME01,SME02},
which is an effective field theory irrelative to the ultimately underlying theory.
Obviously, the spacetime background is still Minkowskian in the SME.
However, Minkowski spacetime should be amended together with Lagrangian of particles and fields
if the Lorentz symmetry is violated (no matter spontaneously or not).

Recently, Kostelecky \cite{Kostelecky_Finsler} proposed that
the SME is closely related to Finsler spacetime which is intrinsically anisotropic.
The coupling constants in the sLIV terms could be related to certain fixed preferred directions in the Finsler structure.
The most fundamental reason is that Finsler geometry \cite{Book by Rund,Book by Bao,Book by Shen}
gets rid of the quadratic restriction on the spacetime structure
such that the Finsler metric depends on directions of the spacetime.
In addition, the isometric transformations reveal that non--Riemannian Finsler spacetime possesses fewer symmetries than
Riemann spacetime \cite{Finsler isometry by Wang,Finsler isometry by Rutz,Finsler isometry,Finsler isometry LiCM}.
These characters imply that Finsler spacetime is intrinsically anisotropic.

Einstein's special relativity resides in a flat Riemann spacetime, namely Minkowski spacetime.
Similarly, the special relativity with LIV effects may reside in a flat Finsler spacetime.
In fact, it is found that the SME--related Finsler spacetime is indeed flat in the sense of Finsler geometry \cite{Kostelecky_Finsler}.
For instance, the simplest SME model, with only one nonvanishing coupling constant \(a_{\mu}\) in the sLIV terms,
leads to a flat spacetime of Randers--Finsler geometry \cite{Randers}.
Actually, the flat Finsler spacetime is called locally Minkowski spacetime \cite{Book by Bao},
which could be viewed as a generalization of Minkowski spacetime.
In addition, doubly special relativity (DSR) \cite{DSR01,DSR02,DSR03,DSR04,DSR05} was found to be incorporated into Finsler spacetime
\cite{DSR in Finsler}, as well as very special relativity (VSR) \cite{VSR,VSR in Finsler}.

As the LIV corresponds to new spacetime, it is valuable to investigate physics compatible with the LIV effects.
In this paper, we try to set up equations of motion for the electromagnetic field in locally Minkowski spacetime.
A Lagrangian is proposed for the electromagnetic field in such a spacetime.
The LIV effects are induced into the Lagrangian in a natural way.
The amended Maxwell equations are obtained via the variation of action.
A formal plane wave solution is obtained for the electromagnetic wave.
The dispersion relation is modified for the electromagnetic wave.
We also study the electromagnetic field at the first order of LIV effects.
We compare these perturbative results with those in the SME framework.
Relations and differences are discussed between Finsler spacetime and the SME.
An interpretation is proposed for the LIV effects as influence of an anisotropic media.
In addition, a constraint on the spacetime anisotropy could be obtained from astrophysical observations on gamma-ray bursts (GRBs).

The rest of the paper is arranged as follows.
In section 2, we briefly discuss the spacetime in Finsler spacetime, especially the locally Minkowski spacetime.
We propose an electromagnetic field model in locally Minkowksi spacetime in section 3.
In section 4, we study this model at the first order of LIV effects and compare it with the SME.
The anisotropic media is invoked to interpret the LIV effects on the electromagnetic field.
In section 5, a constraint on the LIV effects is obtained from the \emph{Fermi}--observations of GRBs
in a specific locally Minkowski metric.
Conclusions and remarks are listed in section 6.

\section{2. Spacetime anisotropy}
\label{Spacetime anisotropy}
Finsler spacetime is defined on the tangent bundle \(TM:=\bigcup_{x\in M}T_{x}M\) instead of the manifold \(M\).
Each element of \(TM\) is denoted by \((x,y)\), where \(x\in M\) and \(y\in T_{x}M\).
Finsler geometry originates from the integral of the form \cite{Book by Rund,Book by Bao,Book by Shen}
\begin{equation}
\label{Finsler geometry}
\int^b_a F\left(x, y\right)d\tau\ ,
\end{equation}
where \(x\) denotes a position and \(y:=dx/d\tau\) denotes a so--called 4--velocity.
The integrand \(F(x,y)\) is called a Finsler structure, which is a smooth, positive and positively 1--homogeneous function
defined on the slit tangent bundle \(TM\backslash \{0\}\).
The positive 1--homogeneity denotes the character \(F(x,\lambda y)=\lambda F(x,y)\) for all \(\lambda>0\).
The Finsler metric is defined as
\begin{equation}
\label{Finsler metric}
g_{\mu\nu}(x,y):=\frac{\partial}{\partial y^\mu}\frac{\partial}{\partial y^\nu}\left(\frac{1}{2}F^2\right)\ .
\end{equation}
Together with its inverse tensor, it is used for raising and lowering indices of tensors.
Note that the Finsler metric becomes Riemannian if it does not depend on \(y\).

A Finsler spacetime (\(M\),\(F\)) is called locally Minkowski spacetime \cite{Book by Bao}
if there is no dependence on \(x\) for Finsler structure \(F\), namely \(F=F(y)\).
Therefore, the Finsler metric \(g_{\mu\nu}\) only depends on \(y\) according to (\ref{Finsler metric}).
In such a spacetime, connections and curvatures vanish.
Therefore, it is flat and maximally symmetric \cite{Finsler isometry LiCM,constant flag curvature}.
The vanishment of the connections implies that a free particle follows a straight line.
It also implies that locally Minkowski spacetime belongs to Berwald spacetime \cite{Book by Bao}.
All tangent spaces of Berwald spacetime are linearly isomorphic to one common Minkowski--normed linear space.
Physically, this character implies that the laws of physics are common at each position in such a spacetime.

In Finsler spacetime, 4--velocity of a free particle is given by the Finsler geodesic equation \cite{Book by Bao}
\begin{equation}
\label{geodesic}
\frac{d^{2} x^{\mu}}{d\tau^{2}}+\Gamma^{\mu}_{\rho\sigma}(x,\frac{dx}{d\tau})\frac{d x^{\rho}}{d \tau}\frac{d x^{\sigma}}{d \tau}=0\ ,
\end{equation}
where \(\Gamma\) denotes the connection.
The Finsler geodesic originates from variation of an integral of the Finsler line element of the form (\ref{Finsler geometry}).
In locally Minkowski spacetime, the connections vanish, particularly.
The Finsler geodesic equation (\ref{geodesic}) becomes
\begin{equation}
\label{geodesic equation}
\frac{d^2 x^{\mu}}{d\tau^{2}}=0\ .
\end{equation}
Its solution gives a constant vector to \(y\), which means that \(y\) is independent on \(x\).
In this paper, \(y\) denotes 4--velocity of a free photon along the Finsler geodesic.
For a charged particle, such as electron, it would interact with the electromagnetic field.
The Finsler geodesic equation should be modified.
An extra term related to electromagnetic force \(F^{\mu}(x)\) should be added to the right hand side (r.h.s.) of the Finsler geodesic equation.
The velocity of the charged particle is given by the solution of the modified geodesic equation.
Thus, it depends on \(x\).

\section{3. Electromagnetic field in locally Minkowski spacetime}
\label{Electromagnetic field in locally Minkowski spacetime}
An advantage of studying the LIV in Finsler spacetime is that the principle of relativity is preserved automatically.
As in Minkowski spacetime, we define the 4--potential 1--form of electromagnetic field in locally Minkowski spacetime
\begin{equation}
\label{1--form}
A:=A_{\mu}(x)dx^{\mu}\ .
\end{equation}
It preserves the internal U(1) gauge symmetry.
The electromagnetic 4--potential is chosen as such a form that its 2--form excludes the terms \(dx^{\mu}\wedge \delta y^{\nu}\) and \(\delta y^{\mu}\wedge \delta y^{\nu}\) whose physical meaning is unclear.
Therefore, the field strength 2--form is given by
\begin{equation}
F:=dA=\frac{1}{2!}F_{\mu\nu}(x)dx^{\mu}\wedge dx^{\nu}\ ,
\end{equation}
where
\begin{equation}
\label{2--form}
F_{\mu\nu}=\frac{\partial A_{\nu}}{\partial x^{\mu}}-\frac{\partial A_{\mu}}{\partial x^{\nu}}\
\end{equation}
is invariant under the U(1) gauge group.

One of the Maxwell equations is given by the Bianchi identity \(dF=0\),
\begin{equation}
\label{fisrt Maxwell's equation}
\frac{\partial F_{\mu\nu}}{\partial x^{\lambda}}+\frac{\partial F_{\nu\lambda}}{\partial x^{\mu}}+\frac{\partial F_{\lambda\mu}}{\partial x^{\nu}}=0\ .
\end{equation}
It is similar to the one in Minkowski spacetime.
In addition, the contravariant field strength \(F^{\mu\nu}\) is given via raising the indices of the covariant 2--form (\ref{2--form})
by the Finsler metric \(g^{\mu\nu}(y)\) of locally Minkowski spacetime, namely \(F^{\mu\nu}=g^{\mu\sigma}g^{\nu\lambda}F_{\sigma\lambda}\).
In this way, the covariant character is preserved in locally Mikowski spacetime.

We follow the form of the Lagrangian for the electromagnetic field
but replace the spacetime metric \(\eta_{\mu\nu}\) by the Finsler metric \(g_{\mu\nu}(y)\) \cite{Stueckelberg method01,Stueckelberg method02,Massive photons__Stueckelberg method}.
In this way, the Lagrangian could reduce back to the one in Minkowski spacetime
when locally Minkowski spacetime reduces into Minkowski spacetime.
The Lagrangian takes the form
\begin{equation}
\label{Lagrangian}
L=-\frac{1}{4}F_{\mu\nu}F^{\mu\nu}\ .
\end{equation}
The LIV effects are introduced via contracting the spacetime indices by the Finsler metric.
The Lagrangian is invariant under coordinate transformations.

In locally Minkowski spacetime, an orthogonal base is given by \(\{\frac{\partial}{\partial x^{\mu}}\}\) and its dual base is \(\{dx^{\mu}\}\).
The action of the electromagnetic field could be given by
\begin{equation}
\label{action 1}
I=\int \left(-\frac{1}{4}F_{\mu\nu}F^{\mu\nu}\right) d\Omega\ ,
\end{equation}
where \(d\Omega=\sqrt{-\det{g_{\mu\nu}(y)}}d^{4}x\) denotes the invariant volume element at each position \(x\).
The variation of action (\ref{action 1}) with respect to \(A_{\mu}\) is given by
\begin{equation}
\label{variation of first term}
\int \left(\frac{\partial L}{\partial A^{\mu}}-\frac{\partial}{\partial x^{\sigma}}\frac{\partial L}{\partial \left(\frac{\partial A^{\mu}}{\partial x^{\sigma}}\right)}\right)\delta A^{\mu}d\Omega=0\ .
\end{equation}
It gives the familiar Euler--Lagrangian equation
\begin{equation}
\label{Euler--Lagrangian equation}
\frac{\partial L}{\partial A^{\mu}}-\frac{\partial}{\partial x^{\sigma}}\frac{\partial L}{\partial \left(\frac{\partial A^{\mu}}{\partial x^{\sigma}}\right)}=0\ .
\end{equation}
The Euler--Lagrange equation supplements the Maxwell equations
\begin{equation}
\label{second Maxwell's equation}
g^{\mu\nu}\frac{\partial F_{\mu\sigma}}{\partial x^{\nu}}=0\ .
\end{equation}
The equations (\ref{fisrt Maxwell's equation}) and (\ref{second Maxwell's equation})
form a complete set of equations of motion for the electromagnetic field in locally Minkowski spacetime.

The Maxwell's equation (\ref{second Maxwell's equation}) could be rewritten in terms of \(A_{\sigma}\) as
\begin{equation}
\label{electromagnetic wave equation}
g^{\mu\nu}\frac{\partial^{2}A_{\sigma}}{\partial x^{\mu}\partial x^{\nu}}=0\ ,
\end{equation}
under the Lorentz gauge
\begin{equation}
\label{Lorentz gauge}
g^{\mu\nu}\frac{\partial A_{\mu}}{\partial x^{\nu}}=0\ .
\end{equation}
The above equation is the so--called electromagnetic wave equation.
It has a formal plane wave solution
\begin{equation}
\label{plane wave solution}
A_{\sigma}\propto \epsilon_{\sigma}e^{-ik_{\mu}x^{\mu}}=\epsilon_{\sigma}e^{-ig_{\mu\nu}k^{\mu}x^{\nu}}\ ,
\end{equation}
where \(\epsilon_{\sigma}\) denotes a polarization and \(k^{\mu}\) denotes a wavevector of the electromagnetic plane wave.
By substituting (\ref{plane wave solution}) into (\ref{electromagnetic wave equation}),
we obtain a dispersion relation for the electromagnetic plane wave
\begin{equation}
\label{dispersion relation of light}
k_{\mu}k^{\mu}=g_{\mu\nu}k^{\mu}k^{\nu}=0\ .
\end{equation}
Its form is as similar as the one in the Lorentz invariant electrodynamics.
However, it is modified by the Finsler metric \(g_{\mu\nu}\) since the contraction of spacetime indices is implicated via this metric.

\section{4. Lorentz invariance violation}
\label{Lorentz invariance violation}
Observations do not show signals of the LIV effects at the present attainable energy scales \cite{Data tables for Lorentz and CPT violation}.
This fact implies that the LIV effects should be very tiny.
We could extract the LIV effects by expanding the Finsler metric into
\begin{equation}
\label{Finsler metric series}
g^{\mu\nu}(y)=\eta^{\mu\nu}+h^{\mu\nu}(y)\ .
\end{equation}
In this way, the first--order LIV effects are extracted and characterized completely by \(h_{\mu\nu}\).

At the leading order, the Lagrangian (\ref{Lagrangian}) of the electromagnetic field could be expanded into
\begin{eqnarray}
L&=&-\frac{1}{4}\eta^{\mu\rho}\eta^{\nu\sigma}F_{\mu\nu}F_{\rho\sigma}
-\frac{1}{2}\eta^{\mu\rho}h^{\nu\sigma}F_{\mu\nu}F_{\rho\sigma}\ ,\\
&:=&L_{LI}+L_{LIV}\ ,
\end{eqnarray}
where \(L_{LI}=-\frac{1}{4}\eta^{\mu\rho}\eta^{\nu\sigma}F_{\mu\nu}F_{\rho\sigma}\) denotes the Lorentz invariant term
while the LIV term is given as
\begin{equation}
\label{sLIV in locally Minkowski spacetime}
L_{LIV}=-\frac{1}{8}\left(\eta^{\mu\rho}h^{\nu\sigma}-\eta^{\nu\rho}h^{\mu\sigma}-\eta^{\mu\sigma}h^{\nu\rho}+\eta^{\nu\sigma}h^{\mu\rho}\right)
F_{\mu\nu}F_{\rho\sigma}\ .
\end{equation}
In the above equation, we have anti--symmetrized the indices \(\mu\nu\) and \(\rho\sigma\).
In the SME framework, meanwhile, the CPT--even sLIV term in the Lagrangian of the electromagnetic field is given by \cite{SME02}
\begin{equation}
\label{sLIV in SME}
L_{SME}=-\frac{1}{4}k^{\mu\nu\rho\sigma}F_{\mu\nu}F_{\rho\sigma}\ ,
\end{equation}
where \(k^{\mu\rho\nu\sigma}\) denotes a dimensionless parameter which characterizes the level of the sLIV effects.
In the SME, the parameter \(k^{\mu\rho\nu\sigma}\) is given by hand.

In locally Minkowski spacetime, however, we could relate this parameter to deformation parameter \(h^{\mu\nu}\) of the spacetime
from Minkowski to locally Minkowski.
By comparing (\ref{sLIV in locally Minkowski spacetime}) and (\ref{sLIV in SME}),
we obtain a relation
\begin{equation}
\label{SME_Finsler}
k^{\mu\nu\rho\sigma}=\frac{1}{2}\left(\eta^{\mu\rho}h^{\nu\sigma}-\eta^{\nu\rho}h^{\mu\sigma}-\eta^{\mu\sigma}h^{\nu\rho}
+\eta^{\nu\sigma}h^{\mu\rho}\right)\ .
\end{equation}
In the SME, the parameter \(k\) is a constant since the energy and momentum are conserved \cite{SME02}.
In locally Minkowski spacetime, the geodesic equation (\ref{geodesic equation}) of photon gives \(y\) a constant vector along the geodesic.
Thus, \(h\) is a constant and the r.h.s. of equation (\ref{SME_Finsler}) is also a constant.
These inferences mean that the LIV electromagnetic field model obtained in locally Minkowski spacetime
is compatible with the perturbative results in the SME.
In addition, there are ten independent components for \(h^{\mu\nu}\) while nineteen for \(k^{\mu\nu\rho\sigma}\) \cite{SME02}.
Only components of the form \(k^{\mu\nu\mu\sigma}\) are possibly nonvanishing in locally Minkowski spacetime.
Furthermore, the birefringence of light will not emerge at the leading order in this Finsler model of electromagnetic field.
The reason is that all Weyl components of \(k^{\mu\nu\rho\sigma}\) vanish
at the leading order \cite{Electrodynamics with Lorentz-violating operators of arbitrary dimension}.
These predictions distinguish the electromagnetic field model in locally Minkowski spacetime from the SME--based one.

As \(L_{SME}\) does in the SME, the LIV term \(L_{LIV}\) also denotes the Lorentz--violating interactions
at first order for the electromagnetic field in locally Minkowski spacetime.
Traditionally, the observations on the electromagnetic field give rise to the most stringent tests of the Lorentz symmetry.
An incomplete list includes:
the LIV could lead to the anisotropy of the speed of light which is tested by the Michelson--Morley experiment
\cite{Michelson--Morley01,Michelson--Morley02,Michelson--Morley03,Michelson--Morley04,Michelson--Morley05};
there is an atomic clock experiment named as the Hughes--Drever experiment \cite{Hughes--Drever exp01,Hughes--Drever exp02}
which is used to test the variation of the SME coefficients with the movement of the Earth;
the observations from distant galaxies give severe limits on the birefringence of light \cite{Birefiringence01,Birefiringence02}, and etc.
For a more detailed summarization on the observations of the (s)LIV effects, see for example
citations \cite{Data tables for Lorentz and CPT violation,Overview of the SME by Bluhm} and references therein.

The Maxwell equations (\ref{fisrt Maxwell's equation}) and (\ref{second Maxwell's equation}) could be rewritten as
\begin{eqnarray}
\label{Maxwell equations in series}
\frac{\partial F_{\mu\nu}}{\partial x^{\lambda}}+\frac{\partial F_{\nu\lambda}}{\partial x^{\mu}}+\frac{\partial F_{\lambda\mu}}{\partial x^{\nu}}&=&0\ ,\\
\eta^{\mu\nu}\frac{\partial F_{\mu\sigma}}{\partial x^{\nu}}+h^{\mu\nu}\frac{\partial F_{\mu\sigma}}{\partial x^{\nu}}&=&0\ .
\end{eqnarray}
The second equation includes the LIV effects while the first one is not related to dynamics.
The electromagnetic wave equation (\ref{electromagnetic wave equation}) could be expanded as
\begin{equation}
\label{electromagnetic wave in series}
\eta^{\mu\nu}\frac{\partial^{2}A_{\sigma}}{\partial x^{\mu}\partial x^{\nu}}+h^{\mu\nu}\frac{\partial^{2}A_{\sigma}}{\partial x^{\mu}\partial x^{\nu}}=0\ ,
\end{equation}
where the first term denotes the electromagnetic wave equation in Minkowski spacetime
and the second term denotes the terms related to the LIV effects.

There is a solution for this wave equation (\ref{electromagnetic wave in series}) at first order
\begin{equation}
A_{\mu}=A_{0\mu}+A_{1\mu}\ ,
\end{equation}
where \(A_{0\mu}\) and \(A_{1\mu}\) denote the zero--order solution and the first--order solution, respectively.
The zero--order solution \(A_{0\mu}\) satisfies the electromagnetic wave equation in Minkowski spacetime, namely
\begin{equation}
\eta^{\mu\nu}\frac{\partial^{2}A_{0\sigma}}{\partial x^{\mu}\partial x^{\nu}}=0\ .
\end{equation}
It has a plane wave solution \(A_{0\sigma}\propto\epsilon_{\sigma} e^{-\eta_{\mu\nu}k^{\mu}x^{\nu}}\).
Therefore, we could obtain an equation for the first--order solution \(A_{1\mu}\) as
\begin{eqnarray}
\eta^{\mu\nu}\frac{\partial^{2}A_{1\sigma}}{\partial x^{\mu}\partial x^{\nu}}&=&-h^{\mu\nu}\frac{\partial^{2}A_{0\sigma}}{\partial x^{\mu}\partial x^{\nu}}\nonumber\\
&=& h^{\mu\nu}\eta_{\mu\rho}\eta_{\nu\kappa}k^{\rho}k^{\kappa}A_{0\sigma}\nonumber\\
&=& h_{\mu\nu}k^{\mu}k^{\nu}A_{0\sigma}\ ,
\end{eqnarray}
where we have contracted indices with \(\eta\) in the third equal.
The r.h.s. of the above equation behaves like a source of the electromagnetic field,
which could be viewed as influence from a slightly anisotropic media on the electromagnetic wave.
Furthermore, the dispersion relation (\ref{dispersion relation of light}) could be expanded into
\begin{equation}
\label{dispersion relation in series}
\eta_{\mu\nu}k^{\mu}k^{\nu}=-h_{\mu\nu}k^{\mu}k^{\nu}\ .
\end{equation}
It is also called the lightcone.
The lightcone is enlarged if the r.h.s. of the above equation is negative,
while narrowed if the r.h.s. of the above equation is positive.
It depends on concrete characters of the LIV effects \(h_{\mu\nu}\).
The spatial speed of light could be superluminal if the lightcone is enlarged
while it is subluminal if the lightcone is narrowed, and vice versa \cite{FSR01,FSR02,FSR03}.
In addition, the speed of light could depend on its direction since there could be of direction--dependence for \(h_{\mu\nu}\).
These could be tested by observations, such as the Michelson--Morley experiment.

\section{5. Constraint from Gamma-Ray Bursts}
\label{Constraint from Gamma-Ray Bursts}
In this section, a specific locally Minkowksi metric is postulated and investigated.
The speed of light is obtained to be subluminal and the lightcone is found to be squeezed.
In addition, a constraint on the level of the LIV effects is gained from the \emph{Fermi}--observations of GRBs.

To discuss detailed predictions on the LIV effects, we postulate the locally Minkowski spacetime as
\begin{equation}
\label{specific locally Minkowski metric}
g^{\mu\nu}=\rm{diag}(1+ay^{0},-1,-1,-1)\ ,
\end{equation}
where \(|ay^{0}|\ll 1\) is assumed and \(a\) is positive.
\(F\) has been normalized \(F(\tilde{y})=0\), and \(y^{\mu}=\tilde{y}^{\mu}/F(\tilde{y})\).
The 4--velocity of a particle is related to 4--momentum of this particle.
Thus, \(y\) could be characterized by \(k\).
In the simplest case, \(y\) is a linear function of \(k\), namely \(y\propto k\) as similar as that in quantum mechanics.
In this way, the metric could be written as
\begin{equation}
\label{specific locally Minkowski metric}
g^{\mu\nu}=\rm{diag}\left(1+\frac{k^{0}}{M},-1,-1,-1\right)\ ,
\end{equation}
where the constant \(M\) is a high--energy scale into which \(a\) has been absorbed.
The perturbative metric deviation is given by
\begin{equation}
\label{metric perturbation in specific locally Minkowski}
h^{00}=-h_{00}=\frac{k^{0}}{M}\ ,
\end{equation}
and other components vanish.
The energy scale \(M\) implies a scale for possible occurrence of the LIV effects.
Meanwhile, it reveals that the LIV effects are suppressed severely by this scale.
Therefore, the LIV effects are expected to be most possibly observed in the ultra--high energy physics, such as the Planck scale.

With the spatially isotropic metric (\ref{specific locally Minkowski metric}),
the electromagnetic wave equation (\ref{electromagnetic wave equation}) (or (\ref{electromagnetic wave in series})) becomes
\begin{equation}
\label{electromagnetic wave equation in specific}
\left[\left(1+\frac{k^{0}}{M}\right)\frac{\partial^{2}}{\partial t^{2}}-\nabla^{2}\right]A_{\sigma}=0\ ,
\end{equation}
where \(\nabla\) denotes the 3D divergence.
Comparing this equation with that in the Lorentz--invariant electrodynamics,
we obtain the speed of light as
\begin{equation}
\label{speed of light}
c=\left(1+\frac{k^{0}}{M}\right)^{-\frac{1}{2}}\approx1-\frac{k^{0}}{2M}\ .
\end{equation}
It implies that a photon with energy \(k^{0}>0\) would propagate subluminally in such a spacetime.
Meanwhile, higher the photon energy is, slower it propagates.
On the other hand, the dispersion relation (\ref{dispersion relation of light}) (or (\ref{dispersion relation in series})) becomes
\begin{equation}
\eta_{\mu\nu}k^{\mu}k^{\nu}=\frac{k^{0}}{M}(k^{0})^{2}>0\ .
\end{equation}
It implies that the lightcone is squeezed.
Higher the photon energy is, more severely its lightcone is squeezed.
These are consistent with the prediction that the speed of light (\ref{speed of light}) is ``subluminal''.

The above predictions could be tested by the astrophysical observations on GRBs.
The reason is that the above LIV effects could be accumulated after photons traveling a cosmological distance.
The \emph{Fermi} satellite has observed several GRBs with photon energy larger than \(100~\rm{MeV}\) in recent years.
It has been shown that \(\rm{GeV}\) photons arrive several seconds later
than \(\rm{MeV}\) photons \cite{GRB080916C,GRB090902B,limit on variation of the speed of light,GRB090926A}.
The observed time lag for two photons with energy \(k^{0}_{high}\) and \(k^{0}_{low}\)
consists of two parts \cite{A unified constraint on the Lorentz invariance violation from GRBs}
\begin{equation}
\label{t obs}
\Delta t_{obs}=\Delta t_{LIV}+\Delta t_{int}\ ,
\end{equation}
where \(\Delta t_{int}\) denotes the intrinsic emission time delay,
and \(\Delta t_{LIV}\) represents the flying time difference induced by the LIV effects.
According to the magnetic jet model \cite{Magnetic jet model for GRBs},  \(\Delta t_{int}\) could be evaluated.
In such a model, photons with energy less than \(10~\rm{MeV}\) can escape
when the jet radius is beyond the Thomson photosphere radius, i.e., the optical depth is \(\tau_{T}\sim 1\).
Nevertheless, \(\rm{GeV}\) photons will be converted into electron--positron pairs at this radius,
but can escape later when the pair--production optical depth \(\tau_{\gamma\gamma}(k^{0})\) drops below unity.
One can calculate the time delay \(\Delta t(k^{0})\) for the emissions of \(\rm{GeV}\) and \(100~\rm{MeV}\) photons relative to
\(\rm{MeV}\) photons (see detailed discussions in Ref.\cite{A unified constraint on the Lorentz invariance violation from GRBs}).
The intrinsic time delay is \(\Delta t_{int}=\Delta t(k^{0}_{high})-\Delta t(k^{0}_{low})\).
Therefore, we could obtain the LIV--induced time delay \(\Delta t_{LIV}\) according to (\ref{t obs}), see Table \ref{tab:1}.
\begin{table*}
\begin{center}
\caption{Table taken from the Ref.\cite{A unified constraint on the Lorentz invariance violation from GRBs}. The LIV--induced time delay \(\triangle\)t\(_{LIV}\) and the LIV energy scale 2M derived from the \emph{Fermi}--observations of four GRBs: GRB 080916c, GRB 090510, GRB 090902b and GRB 090926. The observed time lags \(\triangle\)t\(_{obs}\) were collected from Ref.\cite{GRB080916C,GRB090902B,limit on variation of the speed of light,GRB090926A}. The LIV energy scale was shown to be 2M\(\sim10^{20}\) GeV, which is consistent with the Planck energy.}
\label{tab:1}
\begin{tabular}{ccccccc}
\hline\noalign{\smallskip}
GRB & k\(^{0}\)\(_{\rm{low}}\) & k\(^{0}\)\(_{\rm{high}}\) & $\triangle$ t\(_{\rm{obs}}\) & $\triangle$ t\(_{\rm{LIV}}\) & K(z)         & 2M  \\
~   & MeV     & GeV      & s              & s              & s\(\cdot\)GeV   & GeV \\
\noalign{\smallskip}\hline\noalign{\smallskip}
080916c & 100 & 13.22 & 12.94 & 0.24 & 4.50\(\times10^{18}\) & 10.02\(\times10^{19}\) \\
090510 & 100 & 31 & 0.20 & 0.14 & 7.02\(\times10^{18}\) & 9.73\(\times10^{19}\) \\
090902b & 100 & 11.16 & 9.5 & 0.10 & 3.38\(\times10^{18}\) & 9.94\(\times10^{19}\) \\
090926 & 100 & 19.6 & 21.5 & 0.20 & 6.20\(\times10^{18}\) & 9.59\(\times10^{19}\) \\
\noalign{\smallskip}\hline
\end{tabular}
\end{center}
\end{table*}

Consider two photons emitted at the same spacetime point, the arrival time delay between them could be written as
\cite{A unified constraint on the Lorentz invariance violation from GRBs,high energy photons from Fermidetected GRBs,LIV-induced arrival delays of cosmological particles}
\begin{equation}
\label{t LIV}
\Delta t_{LIV}=\frac{\Delta k^{0}}{2M}D(z)\ ,
\end{equation}
where we have used the equation (\ref{speed of light}).
The cosmological distance \(D\) is defined as
\cite{A unified constraint on the Lorentz invariance violation from GRBs,high energy photons from Fermidetected GRBs,LIV-induced arrival delays of cosmological particles}
\begin{equation}
D(z):=H_{0}^{-1}\int_{0}^{z}\frac{(1+z')dz'}{\sqrt{\Omega_{M}(1+z)^{3}+\Omega_{\Lambda}}}\ ,
\end{equation}
where \(H_{0}\approx 72\rm{km\cdot sec^{-1}\cdot Mpc^{-1}}\)denotes the Hubble constant,
\(\Omega_{M}\approx0.3\) and \(\Omega_{\Lambda}\approx0.7\) are densities of matter and cosmological constant, respectively.
In this way, the LIV energy scale is given by
\begin{equation}
\label{LIV energy scale}
2M=\frac{\Delta k^{0}}{\Delta t_{LIV}}D(z)\ .
\end{equation}
To reveal the LIV effects, one depicts the \(\Delta t_{LIV}/(1+z)~\rm{vs.}~K(z)\) plot,
where \(K(z)\) is defined as
\cite{A unified constraint on the Lorentz invariance violation from GRBs,Lorentz violation from cosmological objects}
\begin{equation}
K(z):=\frac{\Delta k^{0}}{1+z}D(z)\ .
\end{equation}
The slope of this plot denotes the inverse of the level of the LIV effects, i.e., \((2M)^{-1}\).

\begin{figure}[h]
\begin{center}
\includegraphics[width=8 cm]{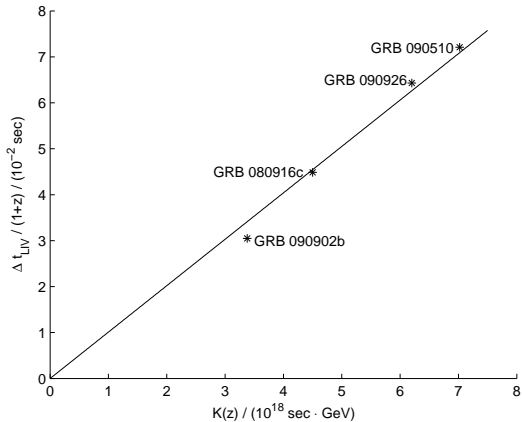}
\caption{Figure taken from Ref.\cite{A unified constraint on the Lorentz invariance violation from GRBs}. The \(\Delta t_{LIV}/(1+z)~\rm{vs.}~K(z)\) plot for the four GRBs observed by the \emph{Fermi} satellite. The slope of the fit line was shown to be \((2M)^{-1}\sim 10^{-20}~\rm{GeV}^{-1}\).}
\label{fig1}
\end{center}
\end{figure}

In Ref.\cite{A unified constraint on the Lorentz invariance violation from GRBs},
Chang {\it et al.} took advantage of the \emph{Fermi}--observations of four GRBs to estimate the level of the LIV effects.
The four GRBs are GRB 080916c \cite{GRB080916C}, GRB 090902b \cite{GRB090902B},
GRB 090510 \cite{limit on variation of the speed of light} and GRB 090926 \cite{GRB090926A}, respectively.
Their LIV--induced time lags \(\Delta t_{LIV}\) and \(K(z)\) were calculated and listed in Table \ref{tab:1}.
Their \(\Delta t_{LIV}/(1+z)~\rm{vs.}~K(z)\) plot was given by Fig.\ref{fig1}.
The slope of the fit line was obtained \((2M)^{-1}\sim 10^{-20}~\rm{GeV}^{-1}\).
Correspondingly, the LIV energy scale was shown to be \(2M\sim 10^{20}~\rm{GeV}\), which is consistent with the Planck energy scale.
Therefore, we would expect to observe the spacetime anisotropy near the planck scale in future astrophysical and cosmological observations.

\section{6. Conclusions and remarks}
\label{Conclusions and remarks}
Conclusions and remark are listed as follows.
Finsler geometry gets rid of the quadratic restriction on the form of the spacetime structure.
It is intrinsically anisotropic.
The SME--related sLIV effects on the classical point--like particles have been related to this kind of intrinsically anisotropic spacetime.
In principle, the laws of physics should be studied in the intrinsically anisotropic spacetime if the Lorentz symmetry is violated.
In this paper, we proposed that locally Minkowski spacetime could be a suitable platform to characterize the possible LIV effects.
The reason is that locally Minkowski spacetime is the flat and maximally symmetric non--Riemannian Finsler spacetime.

We studied the electromagnetic field in the locally Minkowski spacetime.
The Lagrangian with LIV effects was constructed for the electromagnetic field via replacing the spacetime metric with the Finsler metric.
It was found that the obtained Lagrangian is invariant under the coordinate transformations,
which preserves validation of the principle of relativity.
We obtained the Maxwell equations via the Bianchi identity and the variation of action.
We presented a formal plane wave solution. The dispersion relation is modified for the electromagnetic wave.
The lightcone might be enlarged or narrowed, depending on concrete characters of the LIV effects.
The approach proposed in the paper could be generalized straightforward to
study the non--Abelian gauge fields with the LIV effects in locally Minkowski spacetime.

To demonstrate the LIV effects clearly in locally Minkowski spacetime,
we expanded the Lagrangian of the electromagnetic field around the Minkowski background.
The explicit LIV term was especially extracted from the Lagrangian.
It is noteworthy that this LIV term could be reduced back to the sLIV one in the SME formally at first order.
It reveals that our results are compatible with the previous works in the framework of SME.
However, the LIV effects originate in departure from Minkowski spacetime to locally Minkowski spacetime.
There are fewer independent parameters for the LIV effects in locally Minkowski spacetime.
The birefringence of light would not appear in our model, which is consistent with the astronomical observations.
In addition, the LIV influence on the electromagnetic wave was found to behave like a source of the electromagnetic field.
It could be interpreted as influence from a slightly anisotropic media on the electromagnetic field.

To discuss phenomenological predictions on the LIV effects, we investigate a specific locally Minkowski metric.
The electromagnetic wave equation was studied and the light was found to propagate subluminally.
On the other hand, we obtained a squeezed lightcone.
Both characters are consistent with each other.
Another important feature of this metric was that the lightcone becomes more severely squeezed as increase of the photon energy.
These features were tested by the \emph{Fermi}--observations on the GRBs.
The LIV effects accumulate when the light propagates from distant GRBs.
The GeV photons were found to arrive at the Earth later than the MeV photons.
This observation gave a severe constraint on the LIV energy scale, i.e., \(10^{20}~\rm{GeV}\).
We would expect to observe the spacetime anisotropy near this energy scale in future astrophysical and cosmological observations.

\vspace{0.4 cm}

\begin{acknowledgments}
We thank useful discussions with Yunguo Jiang, Ming-Hua Li, Xin Li, Hai-Nan Lin.
The author (S. Wang) thanks useful discussions with Jian-Ping Dai, Dan-Ning Li, and Xiao-Gang Wu.
This work is supported by the National Natural Science Fund of China under Grant No. 11075166.
\end{acknowledgments}


\begin{thebibliography}{999}
\bibitem{DSR01}G. Amelino-Camelia, Phys. Lett. B {\bf 510}, 255 (2001).
\bibitem{DSR02}G. Amelino-Camelia, Int. J. Mod. Phys. D {\bf 11}, 35 (2002).
\bibitem{DSR03}G. Amelino-Camelia, Nature {\bf 418}, 34 (2002).
\bibitem{DSR04}J. Magueijo and L. Smolin, Phys. Rev. Lett. {\bf 88}, 190403 (2002).
\bibitem{DSR05}J. Magueijo and L. Smolin, Phys. Rev. D {\bf 67}, 044017 (2003).
\bibitem{spacetime foam01}J. Alfaro, H. A. Morales-Tecotl and L. F. Urrutia, Phys. Rev. Lett. {\bf 84}, 2318 (2000).
\bibitem{spacetime foam02}J. Alfaro, H. A. Morales-Tecotl and L. F. Urrutia, Phys. Rev. D {\bf 65}, 103509 (2002).
\bibitem{spacetime foam03}D. Sudarsky et al., Phys. Rev. D {\bf 68}, 024010 (2003).
\bibitem{spacetime foam04}J. Bernabeu, N. E. Mavromatos and Sarben Sarkar, Phys. Rev. D {\bf 74}, 045014 (2006).
\bibitem{VSR}A. G. Cohen and S. L. Glashow, Phys. Rev. Lett. {\bf 97}, 021601 (2006).
\bibitem{Pavlopoulos}T. G. Pavlopoulos, Phys. Rev. {\bf 159}, 1106 (1967).
\bibitem{KosteleckyS001}V. A. Kostelecky and S. Samuel, Phys. Rev. D {\bf 39}, 683 (1989).
\bibitem{SME01}D. Colladay and V. A. Kostelecky, Phys. Rev. D {\bf 55}, 6760 (1997).
\bibitem{SME02}D. Colladay and V. A. Kostelecky, Phys. Rev. D {\bf 58}, 116002 (1998).
\bibitem{Kostelecky_Finsler}V. A. Kostelecky, Phys. Lett. B {\bf 701}, 137 (2011).
\bibitem{Book by Rund}H. Rund, {\it The Differential Geometry of Finsler Spaces}, Springer, Berlin, 1959.
\bibitem{Book by Bao}D. Bao, S. S. Chern, and Z. Shen, {\it An Introduction to Riemann--Finsler Geometry},
        Graduate Texts in Mathmatics {\bf 200}, Springer, New York, 2000.
\bibitem{Book by Shen}Z. Shen, {\it Lectures on Finsler Geometry}, World Scientific, Singapore, 2001.
\bibitem{Finsler isometry by Wang}H. C. Wang, J. London Math. Soc. {\bf s1-22} (1), 5 (1947), DOI:10.1112/jlms/s1-22.1.5.
\bibitem{Finsler isometry by Rutz}S. F. Rutz, Contemp. Math. {\bf 169}, 289 (1996).
\bibitem{Finsler isometry}S. Deng and Z. Hou, Pac. J. Math. {\bf 207}, 1 (2002).
\bibitem{Finsler isometry LiCM}X. Li and Z. Chang, arXiv:1010.2020 [gr-qc].
\bibitem{Randers}G. Randers, Phys, Rev. {\bf 59}, 195--199(1941).
\bibitem{DSR in Finsler}F. Girelli, S. Liberati and L. Sindoni, Phys. Rev. D {\bf 75}, 064015 (2007).
\bibitem{VSR in Finsler}G.W. Gibbons, J. Gomis and C.N. Pope, Phys. Rev. D {\bf 76}, 081701 (2007).
\bibitem{constant flag curvature}B. Li and Z. Shen, International Journal of Mathematics {\bf 18}, 1 (2007).
\bibitem{Stueckelberg method01}E. C. G. Stueckelberg, Helv. Phys. Acta {\bf 11}, 225 (1938).
\bibitem{Stueckelberg method02}E. C. G. Stueckelberg, Helv. Phys. Acta {\bf 11}, 299 (1938).
\bibitem{Massive photons__Stueckelberg method}M. Cambiaso, R. Lehnert and R. Potting, Phys. Rev. D {\bf 85}, 085023 (2012).
\bibitem{Data tables for Lorentz and CPT violation}V. A. Kostelecky and N. Russell, Rev. Mod. Phys. {\bf 83}, 11 (2011).
\bibitem{Electrodynamics with Lorentz-violating operators of arbitrary dimension}A. Kostelecky and M. Mewes, Phys. Rev. D {\bf 80}, 015020 (2009).
\bibitem{Michelson--Morley01}J. Lipa et al., Phys. Rev. Lett. {\bf 90}, 060403 (2003).
\bibitem{Michelson--Morley02}H. Muller, S. Herrmann, C. Braxmaier, S. Schiller and A. Peters, Phys. Rev. Lett. {\bf 91}, 020401 (2003).
\bibitem{Michelson--Morley03}P. Wolf et al., Gen. Rel. Grav. {\bf 36}, 2351 (2004).
\bibitem{Michelson--Morley04}P. Wolf et al., Phys. Rev. D {\bf 70}, 051902 (2004).
\bibitem{Michelson--Morley05}M.E. Tobar et al., Phys. Rev. D {\bf 71}, 025004 (2005).
\bibitem{Hughes--Drever exp01}V.W. Hughes, H.G. Robinson, and V. Beltran-Lopez, Phys. Rev. Lett. {\bf 4}, 342 (1960).
\bibitem{Hughes--Drever exp02}R.W.P. Drever, Philos. Mag. {\bf 6}, 683 (1961).
\bibitem{Birefiringence01}V.A. Kostelecky and M. Mewes, Phys. Rev. Lett. {\bf 87}, 251304 (2001).
\bibitem{Birefiringence02}V.A. Kostelecky and M. Mewes, Phys. Rev. D {\bf 66}, 056005 (2002).
\bibitem{Overview of the SME by Bluhm}R. Bluhm, Lec. Notes Phys. {\bf 702}, 191 (2006).
\bibitem{FSR01}Z. Chang, X. Li and S. Wang, Mod. Phys. Lett. A {\bf 27}, 1250058 (2012).
\bibitem{FSR02}Z. Chang, X. Li and S. Wang, Phys. Lett. B {\bf 710}, 430 (2012).
\bibitem{FSR03}Z. Chang, X. Li and S. Wang, arXiv:1201.1368.
\bibitem{GRB080916C}A. A. Abdo {\it et al.}, Science {\bf 323}, 1688 (2009).
\bibitem{GRB090902B}A. A. Abdo {\it et al.}, Astrophys. J. {\bf 706}, L138 (2009).
\bibitem{limit on variation of the speed of light}A. A. Abdo {\it et al.}, Nature {\bf 462}, 331 (2009).
\bibitem{GRB090926A}M. Ackermann {\it et al.}, Astrophys. J. {\bf 729}, 114 (2011).
\bibitem{A unified constraint on the Lorentz invariance violation from GRBs}Z. Chang, Y. Jiang and H. -N. Lin, arXiv:1201.3413, DOI:10.1016/j.astropartphys.2012.04.006, acceptted by Astropart. Phys..
\bibitem{Magnetic jet model for GRBs}Z. Bosnjak and P. Kumar, Mon. Not. Roy. Astron. Soc. (2012), DOI:10.1111/j.1745-3933.2011.01202.x.
\bibitem{high energy photons from Fermidetected GRBs}R. J. Nemiroff, J. Holmes and R. Connolly, arXiv:1109.5191.
\bibitem{LIV-induced arrival delays of cosmological particles}U. Jacob and T. Piran, JCAP {\bf 01}, 031 (2008).
\bibitem{Lorentz violation from cosmological objects}L. Shao, Z. Xiao and B. -Q. Ma, Astropart. Phys. {\bf 33}, 312 (2010).

\end{thebibliography}
\end{document}